
\input phyzzx.tex
\tolerance=1000
\nopubblock
\vsize = 22 true cm
\hsize = 17 true cm
\hoffset = 0 true cm
\doublespace
\def\ick{\eqalignno}
\twelvepoint

\line{\hfill UdeM-LPN-TH-94-200}
\line{\hfill CRM-2195}
\line{\hfill hep-th/9406187}


\titlepage
\title{Effective potential for nonrelativistic non-Abelian
Chern-Simons matter system in constant background fields$^*$}
\footnote{*}{This work is supported in part by funds provided by the
Natural Sciences and Engineering
Research Council of Canada and the Fonds pour la Formation de Chercheurs et
l'Aide \`a la Recherche.}

\author{Didier Caenepeel$^1$ and Martin Leblanc$^{1,2}$}

\vskip 20 pt

{\baselineskip=12pt

\centerline{\bf \it $^1$ Laboratoire de Physique Nucl\'eaire,}
\baselineskip=12pt
\centerline{\bf \it $^2$ Centre de Recherches Math\'ematiques,}
\vskip 4 pt
\centerline {\bf \it Universit\'e de Montr\'eal }
\centerline {\bf \it Case postale 6128, succ. centre-ville}
\centerline {\bf \it Montr\'eal, (Qc), Canada}
\centerline {\bf \it H3C-3J7 }}

\doublespace

\vskip 40 pt

\abstract{We present the effective potential for nonrelativistic
matter coupled to
non-Abelian Chern-Simons gauge fields. We perform the calculation using a
functional method in constant background fields to satisfy Gauss's law and to
simplify the computation. Both the quantum gauge and matter fields are
integrated over. The gauge fixing is achieved with an $R_\xi$-gauge in the
$\xi\to 0$ limit. Divergences appearing in the matter sector are regulated
via operator regularization. We find no correction to the Chern-Simons
coupling constant and the system experiences conformal symmetry breaking to
one-loop order except at the known value of self-duality. These results agree
with previous analysis of the non-Abelian Aharonov-Bohm scattering.}

\vfill\eject

\normalspace
\line{\bf 1. Introduction \hfil}
\rm

Chern-Simons theories have been studied in many context in the last decade from
the study of general relativity to condensed matter systems. An important line
of developments occurred when it was shown that classical relativistic charged
scalars minimally coupled to an Abelian Chern-Simons gauge field in (2+1)
spacetime dimensions have vortex (soliton) solutions for self-dual equations
when the coupling constant takes special values in a $\phi^6$-theory [1,2]. The
presence of vortex solutions permits the emergence of new mechanisms for anyons
superconductivity [3]. Evidence has been found showing that the existence of
such systems possessing vortex solutions is due to the presence of an $N=2$
supersymmetry obtained by adding fermion fields in an appropriate way [4,5].

It is more reasonable to think that the physics of superconductors should be at
lower energies and described by a nonrelativistic system. It turns out that the
same statements as above can be made for the corresponding nonrelativistic
field theory. Specifically, by taking the limit $c\rightarrow\infty$ ($c$ being
the speed of light), one obtains a field theory of interacting nonrelativistic
scalar fields minimally coupled to an Abelian Chern-Simons gauge field [6,7].
This theory also contains self-dual vortex (soliton) solutions when the
coupling constant takes a special value [7]. Perhaps more surprisingly in the
nonrelativistic case, the self-duality originates also from $N=2$ supersymmetry
[8].

Much work has already been done in generalizing these ideas to non-Abelian
theories. Relativistic and nonrelativistic models of matter fields
coupled to non-Abelian Chern-Simons field [9-11]
have been studied at the classical level, however, the
relation between them as the limit $c\rightarrow\infty$ has never been analysed
as above. Nevertheless, non-Abelian self-dual solitons exist in the
corresponding nonrelativistic Chern-Simons field theory [11]. Supersymmetric
extensions for the relativistic system have proven to show the same relation
between supersymmetry and self-duality as is the case for the Abelian theories
[12,13] and it could be interesting to see if a supersymmetric extension of
the nonrelativistic non-Abelian Chern-Simons matter system is possible.

The quantization of the above models has been discussed in various context. In
the case of the pure non-Abelian Chern-Simons theory, Pisarski {\it et al} [14]
have shown using a perturbative analysis with dimensional regularization that a
one-loop radiative correction to the Chern-Simons coupling constant $\kappa\to
\kappa + {c_2(G)\over 2}$ (shifted by the casimir of the group) occurs. The
same result was then obtained by Witten [15] with a saddle point quantization
around pure gauge vector potentials. Their calculations were confirmed by using
a modified Pauli-Villars method [16], an $F^2$-type regulator [10] and a
modified operator regularization method for determining phases of determinants
[17]. However, it is possible that this shift of the Chern-Simons coupling
constant be absent if a variant of dimensional regularization [10] or a
BRST-invariant regulator is used [18].

When relativistic matter fields are included, Chen {\it et al.} showed that
infinite renormalization for the matter fields as well as for the Chern-Simons
gauge field is necessary at two loops and therefore that the fields obtain
nontrivial anomalous dimensions. Also, the $\beta$-function for the gauge
coupling constant is zero to two-loop order [10].

In the case when nonrelativistic matter fields are coupled to the Abelian
Chern-Simons field, we know that the theory experiences conformal symmetry
breaking at the quantum level unless the coupling constant takes the self-dual
value and that this result holds up to three loop order [19-22]. Only recently
was a perturbative analysis performed for the scattering of scalars in the
nonrelativistic non-Abelian theory using Feynman's diagrammatic [23]. Again,
the conformal symmetry is restored upon choosing the self-dual point. All these
computations were performed either in the conventional Feynman diagrammatic or
within a functional method.

Our goal in this paper is to complement the above discussion and to compute
the scalar field effective potential of the nonrelativistic non-Abelian
Chern-Simons system with the help of a functional method.

We start with an SU(2) non-Abelian Chern-Simons system action
[${\rm diag}\;\;\eta~=(+,-,-)$]

$$\eqalignno {
S= \int dt d^2{\bf x} \;\;
&\Bigl \{- \kappa \epsilon^{\alpha\beta\gamma}{\rm Tr}
(A_\alpha \partial_\beta A_\gamma + {2\over 3}A_\alpha A_\beta A_\gamma)
+i\phi^\dagger D_t \phi
- {1 \over 2 } |{\bf D}\phi|^2
-{\lambda_{pqrs}\over 4}\phi^\dagger_p\phi^\dagger_q \phi_r \phi_s
\Bigr \} \cr
&
&(1.1) \cr }
$$
where the gauge fields belong to the su(2) Lie algebra
$A_\mu=i{A_\mu^a \tau^a\over 2} $,
and $ D_t = \partial_t + i A_0^a {\tau^a\over 2}$ and
${\bf D}={\bf \nabla }-i{\bf A}^a {\tau^a\over 2}$
are the time and space covariant derivatives respectively. $\phi_p$ is the
two component nonrelativistic scalar field, $p=1,2$. The self-interaction
coupling constants satisfy
$\lambda_{pqrs}= \lambda_{qpsr}$ since the fields are bosonic
and $\lambda^*_{pqrs}=\lambda_{rspq}$ for the Lagrangian to be real.
$\tau^a$ are the Pauli matrices
which satisfy the usual commutation relations
$[{\tau^a\over 2},{\tau^b\over 2}]=\epsilon^{abc}{\tau^c\over 2}$
and trace relation ${\rm Tr} \bigl( {\tau^a\over 2}{\tau^b\over 2}\bigr)
= {1\over 2} \delta^{ab}$. We have omitted the mass parameter since in
nonrelativistic systems, it is always possible to set it equal to unity.
[We will use a vector notation: for instance, in the plane the cross product is
${\bf V}\times {\bf W} = \epsilon^{ij}V^iW^j$, the curl of a vector is
${\bf\nabla}\times {\bf V}=\epsilon^{ij}\partial_iV^j$, the curl of a scalar is
$({\bf\nabla}\times S)^i=\epsilon^{ij}\partial_jS$ and we shall introduce the
notation $\bigl ({\bf A}\times {\bf {\hat z}}\bigr)^i =\epsilon^{ij}A^j$. The
notation $x=(t, {\bf x})$ will also be used unless stated otherwise.]

The action (1.1) enjoys several invariances at the classical level.
It is obvious that the matter part of this action is Galilean invariant
and conformally invariant [6,7].
The presence of the Chern-Simons term as the only kinematical term for
gauge fields does not spoil these two sets of invariances as this term is
topologically invariant {\it i.e.} it is invariant upon any space and
time transformations [9,24]. Nevertheless, the interesting
symmetry is gauge invariance. Let us for a moment forget the self-interacting
part of the matter sector. The matter fields are minimally coupled,
hence this part is gauge invariant. The self-interacting part however is
gauge invariant only for
$$
\lambda_{1111}=2\lambda_{1212}=\lambda_{2222} \equiv \lambda \eqno (1.2)
$$
with the other constants vanishing.
The Chern-Simons term is not invariant against gauge transformation; rather
it changes by total derivatives. Under special circumstances the total
derivatives can be set to zero. However, if we need to consider large
gauge transformations, only the exponential of $i\times {\rm action}$ need
be gauge invariant. In this case, we speak of ``gauge invariant action"
if the quantization condition $4\pi\kappa={\rm integer}$ applies [9].
We will use these assessments in the course of the calculation.

To compute the scalar field effective potential of the action (1.1), we proceed
with the functional method of Jackiw [25], which is a useful way to evaluate
the
effective potential without having to use a classical background field, in
conjunction with the operator regularization method [26,27]. The difference
here with other evaluations of effective potentials is that the
electromagnetic field is linearly related to the matter field through Gauss's
law. The procedure involves shifting the fields present in the action by
constants. Shifting the matter field by a constant implies that the magnetic
field must be constant and consequently, quantal effects could emerge from the
background gauge field towards the scalar field effective potential. In the
Abelian version of this model, the same procedure was used to compute the
effective potential for scalars [22].
However in that case, the magnetic field
$B=\nabla\times {\bf A}(x)= {\rm constant}$ was satisfied by a vector
potential, which depended linearly on $x$. In the non-Abelian case, it is
possible to satisfy Gauss's law with a constant vector
potential [see below].

The paper is organised as follows: In the next section, we set up the
problem and provide the
classical equation of motion to show how we satisfy the ones for
the electromagnetic fields with constant background fields.
We show that although local gauge invariance is lost through such a
choice of background fields, global gauge invariance is retained; we gauge fix
in a Galilean fashion in a gauge reminiscent of the
$R_\xi$-gauge, which is globally gauge invariant.
In the third section, we present the results of the calculation and in the
last section, we summarize the work and conclude .

\vglue 0.4cm
\line{\bf 2. Constant background fields, equation of motion and
gauge invariance. \hfil}
\vglue 0.1cm
The functional evaluation of the
effective potential starts with the definition of a new
shifted action:
$$\eqalignno {
S_{\rm new}&=S\Bigl\{\phi_p(x)=\varphi_p+\pi_p(x);
A^a_\mu (x)=a_\mu^a + Q^a_\mu (x)\Bigr \}\cr
&\quad -S\Bigl \{ \varphi_p, a^a_\mu \Bigr\}
- {\rm terms\;linear\;in\;quantum\;fields} &(2.1)\cr }
$$
where we shift the scalar field and the vector potential by constant fields in
such a way that the classical equations of motion for the electromagnetic
fields are satisfied. It is the action (2.1) that enters the path integral
for the evaluation of the effective potential.

Let us for a moment
look at the classical equation of motion of the action (1.1) to see
how the equation of motion respond to the constant shift.
The gauge covariant classical equation of motion for arbitrary fields are
$$\eqalignno {
B^a&\equiv {\bf \nabla}\times {\bf a}^a + {1\over 2}\epsilon^{abc}
{\bf a}^b\times {\bf a}^c
=-{1\over 2\kappa} \phi^\dagger\tau^a \phi &(2.2a) \cr
{\bf E}^a &\equiv - {\bf \nabla} a_0^a - \partial_t {\bf a}^a
+\epsilon^{abc}a_0^b{\bf a}^c
= {1\over 4\kappa} {\bf J}^a\times {\bf {\hat z}}
&(2.2b) \cr
i(D_t&\phi)_p+{1\over 2} ({\bf D}\cdot{\bf D}\phi)_p
-{1\over 2}\lambda_{pqrs}\phi^*_q\phi_r\phi_s = 0 &(2.2c) \cr }
$$
where the current is given by
${\bf J}^a = {1\over 2i}
\bigl [\phi^\dagger({\tau^a\over 2}) ({\bf D}\phi) -
({\bf D}\phi)^\dagger({\tau^a\over 2}) \phi \bigr ] $
and ${\bf D}$ is
the covariant derivative with respect to the background gauge field.

The equation for the
magnetic field (2.2a) is recognized as Gauss's law.  Since scalar fields are
shifted by constants, Eqs.(2.2) have to be read with
$\phi_p=\varphi_p={\rm constant}$.
To maintain consistency with Gauss's law, we need to
choose a background vector potential ${\bf a}^a$ such that the magnetic field
is
constant throughout the plane. The simplest choice is to take
a constant background vector potential [28]. We can also choose
$a_0^a$ constant with the help of Eq.(2.2b).
Since Eqs.(2.2) with constant
background fields are now globally
gauge covariant, without fear of loosing generality,
we can find an explicit solution to Eqs.(2.2a,b).
If we choose for instance, $\varphi_p=(v,0)$ and if we label the SU(2)
group structure with colors $a=(1,2,3)\equiv (Y,B,R)$ then we find that
${\bf a}_Y=(\sqrt {{v^*v\over 2\kappa}},0)$,
${\bf a}_B=(0,-\sqrt {{v^*v\over 2\kappa}})$,
$a^0_R={v^*v\over 16\kappa}$ and the other components vanishing,
is a solution to Eqs.(2.2a,b). Of course, this particular solution
does not satisfy the equation of
motion for the scalar field, Eq.(2.2c), unless $\varphi_p=0$ or if the
coupling constants satisfy $\lambda=-{5\over 16\kappa}$. We will
use the above solution for $a_\mu^a$ and $\varphi_p$
in the definition of $S_{\rm new}$ in Eq.(2.1) and
extrapolate at the end of the calculation
the form of the effective potential since it must be
globally gauge invariant [see below].

We now turn to a proof that global gauge invariance is retained upon
quantizing this theory around constant background fields.
We follow the discussion of Abbott [29].
Under local gauge transformations
$$\eqalignno {
A^\prime_\mu &= U^{-1}\partial_\mu U + U^{-1}A_\mu U &(2.3a) \cr
\phi^\prime &= U^{-1} \phi &(2.3b)\cr}
$$
or infinitesimally with $U=\exp i\omega^a \tau^a/2$
$$\eqalignno {
\delta A^a_\mu&=\partial_\mu\omega^a-\epsilon^{abc}A_\mu^b \omega^c &(2.4a)\cr
\delta\phi_p&=-i\omega^a(\tau^a/2)_{pq}\phi_q &(2.4b)\cr }
$$
the action of Eq.(1.1) transforms according to
$$
\delta S = (4\pi\kappa)(2\pi)w(U) \eqno (2.5)
$$
where $w(U)$ is the winding number and the usual quantization condition
over $4\pi\kappa~={\rm integer}$ follows if we want
$\exp \{i\times {\rm action}\}$
to be gauge invariant under (large) gauge transformation.

In Jackiw's approach the generating functional is defined as
$$\eqalignno {
Z(\varphi_p; a^a_\mu)&= \int \delta \pi \delta Q \;\;det\bigl [
{\delta G^a\over \delta \omega^b } \bigr ] &(2.6) \cr
\exp i \int dt\;d^2{\bf x}
&\Bigl [ {\cal L}
(\varphi_p+\pi_p; a^a_\mu+Q^a_\mu)+{1\over 2\xi} G_a^\dagger G^a
-{\cal L}(\varphi_p, a^a_\mu)
- {\delta {\cal L}\over \delta A}|_{a,\varphi}\cdot Q
-{\delta {\cal L}\over \delta \phi}|_{a,\varphi}\cdot \pi \Bigr] \cr}
$$
where $\varphi_p$ and $a^a_\mu$ are constants, and
${\delta G^a\over \delta \omega^b } $ is the ghost contribution
and is given by the functional derivative of the gauge-fixing term under
the infinitesimal quantum gauge transformation
$\delta Q^a_\mu~=\partial_\mu\omega^a-\epsilon^{abc}(a_\mu^b+Q_\mu^b)\omega^c$.
Then, just as in the conventional approach, the effective
potential at vanishing external current and vanishing quantum field
argument is
$$
\eqalignno {
V_{\rm eff}[\varphi_p, a^a_\mu]&= {i\over \int d^3x}
\ln Z [\varphi_p, a^a_\mu] &(2.7) \cr}
$$

It remains to choose the background field gauge condition which reveals to be
rather difficult for the problem at hand. The reason
is as follows: when matter is not present, the Chern-Simons theory is defined
without the introduction of a metric. Upon choosing the gauge-fixing
condition the theory could loose its topological character [15]. Indeed,
Witten chose a Lorentz-type family gauges in his derivation of the one-loop
quantum correction to the pure non-Abelian Chern-Simons theory. He found,
however, that the topological property of the action remained unaffected. In
the case where matter is coupled to the Chern-Simons theory, we already have
chosen a metric and we must preserve as many as the symmetry present there. In
our case, we have to preserve the Galilean symmetry. We therefore choose
$$
\eqalignno {
G^a&= \nabla\cdot{\bf Q^a}+{i\over 2}\xi
\pi^\dagger \tau^a \varphi &(2.8)\cr }
$$
and note that the gauge-fixing resembles the $R_\xi$-type gauge-fixing
conditions.

Now, by making the following change of variables for the quantum fields
$$
\eqalignno {
Q_\mu &\to Q^\prime_\mu = U^{-1}Q_\mu U \cr
\pi &\to \pi^\prime = U^{-1}\pi &(2.9) \cr }
$$
where $U$ is a gauge transformation with constant $\omega^a$,
it is easy to show
that $Z[\varphi_p,a^a_\mu]$ and hence the effective potential
are invariant under the constant background gauge transformation
$$
\eqalignno {
a_\mu &\to a^\prime_\mu=U^{-1}a_\mu U \cr
\varphi &\to \varphi^\prime=U^{-1}\varphi &(2.10)\cr }
$$
since each term is invariant. It is interesting to note that in retaining only
global gauge invariance, the gauge-fixing condition becomes simpler since it is
not written in an explicit background gauge covariant form [29]. This will of
course be advantageous in the course of the explicit calculation since it will
enables us to integrate out the gauge and matter fields by performing
determinants as they are now diagonal in Fourier space [see below]. We now turn
to the calculation of the effective potential in the $R_\xi$-gauge.

\vglue 0.4cm
\line{\bf 3. The effective potential. \hfil}
\vglue 0.1cm

We perform the calculation of the scalar field effective potential
following the procedure set up in the previous section. The quadratic part
in quantum fields of the action appearing in Eq.~(2.6) upon using the
gauge-fixing condition of Eq.~(2.8) and
$\varphi_p=(v,0)$, ${\bf a}_Y=(\sqrt {{v^*v\over 2\kappa}},0)$,
${\bf a}_B=(0,-\sqrt {{v^*v\over 2\kappa}})$,
$a^0_R={v^*v\over 16\kappa}$ and the other components vanishing
is
$$
\eqalignno {
S=\int dt \; d^2{\bf x} \; &
\Bigl\{ {\kappa\over 2}(\partial_t{\bf Q}_a)\times {\bf Q}_a
-\kappa Q_a^0 {\bf \nabla}\times {\bf Q}_a
+ {1\over 2\xi}({\bf \nabla}\cdot {\bf Q}_a)^2 - {\rho\over 8}{\bf Q}_a\cdot
{\bf Q}_a\cr
&+i\pi^\dagger(D_t)\pi - {1\over 2}|{\bf D}\pi|^2 + {\cal L}_{\rm S.I}
+{\xi\over 8} (\pi^\dagger\tau^a \varphi)(\varphi^\dagger\tau^a\pi) \cr
&+ R^0Q^0_R + B^0Q^0_B + Y^0Q^0_Y +
{\bf R}\cdot {\bf Q}_R + {\bf B}\cdot {\bf Q}_B +
{\bf Y}\cdot {\bf Q}_Y \Bigr\} &(3.1)\cr }
$$
where $\rho =v^*v$, ${\cal L}_{\rm S.I.}$ stands for the
quadratic self-interacting
part in $\pi$-fields, which
will be treated later, and the currents are given by
$$\eqalignno{
R^0&=[ j_R + \kappa ({\bf a}_B\times {\bf Q}_Y -
{\bf a}_Y \times {\bf Q}_B) ]\cr
B^0&= j_B \cr
Y^0&=j_Y &(3.2) \cr
{\bf R} &= \kappa [ {\bf a}_B Q^0_Y - {\bf a}_Y Q^0_B ] \times {\bf {\hat z}}
  \cr
{\bf B} &= {1\over 2}{\bf a}_B j_R + \kappa a^0_R {\bf Q}_Y
\times {\bf {\hat z}}  \cr
{\bf Y} &= {1\over 2}{\bf a}_Y j_R \cr }
$$
with the useful definition for matter-currents
$$\eqalignno{
j_R &= -{1\over 2}(\pi^*_1 v + v^*\pi_1 ) \cr
j_B &= -{i\over 2}(\pi^*_2 v - v^*\pi_2 ) &(3.3) \cr
j_Y &= -{1\over 2}(\pi^*_2 v + v^*\pi_2 ) \quad .\cr }
$$

We are now ready to proceed with the functional integration in the
$\xi\to 0$ limit and up to include ${\cal O}(v^4)$ contributions
[we refer to ${\cal O}(v^4)$ whenever we have ${\cal O}(\lambda^2)$,
${\cal O}({\lambda\over \kappa})$ or ${\cal O}({1\over \kappa^2})$].
The contribution coming from the ghosts to one-loop order
is given by the determinant of the functional derivative
of the gauge-fixing term Eq.~(2.8) with respect to an infinitesimal
quantum gauge transformation as above Eq.~(2.7) without terms
having quantum fields
$$ \eqalignno {
{\rm det} {\delta G^a\over \delta \omega^b} &=
 {\rm det} \Bigl [-\nabla^2 \delta^{ab} -\epsilon^{acb}{\bf a}_c\cdot\nabla
-{\xi\over 4}(v^*, 0) \tau^b\tau^a {v\choose 0} \Bigr ]\quad . &(3.4) \cr }
$$
This contribution is easily calculated since it factorizes from the
path integral and the determinant is performed on a $3\times 3$ matrix.
The result to ${\cal O}(\xi)$ is
$$\eqalignno {
V_{\rm ghosts}
&= -\tr \ln \Bigl ( 1- {( {\bf a}_B\cdot {\bf p})^2\over {\bf p}^4}
- {( {\bf a}_Y\cdot {\bf p})^2\over {\bf p}^4} \Bigr ) &(3.5) \cr }
$$
where the trace is now taken only on energy/momentum space.

Next, we integrate out the quantum gauge fields by integrating first over
the R-color. The first line in Eq.~(3.1) is diagonal in the
(R,B,Y) colors, however, the last line in Eq.~(3.1) mixes the Q's
with different colors. For instance in the R-sector, the structure of the
exponent in the functional integration
is $-{1\over 2}Q^\mu_R \Delta^{-1}_{\mu\nu}Q^\nu_R +Q^\mu_R \; R_\mu$
where in Fourier space (i$\partial_\mu = p_\mu$)
$$
{\Delta}^{-1}(v ;\omega, {\bf p})=
\pmatrix {0& -m & n \cr
              m  &{\rho\over 4}-{1\over \xi}p^1p^1 &-i\kappa\omega-{1\over\xi}
p^1p^2\cr
          -n &i\kappa\omega-{1\over \xi}p^1p^2 &
{\rho\over 4}-{1\over \xi}p^2p^2 \cr }\quad ,\eqno (3.6)
$$
with $m=ic\kappa p^2 $ and $n=ic\kappa p^1$.
Upon the usual change of variable, one obtains a contribution
to the effective potential of the type
$\ln \det^{-1/2}\Delta^{-1}_{\mu\nu}$ and a modification to the original
action by the amount ${1\over 2}R^\mu \Delta_{\mu\nu}R^\nu$, which
does not contain any $Q^\mu_R$-dependence with
$$
{\Delta}(v ;\omega, {\bf p})= -{1\over c^2\kappa^2 {\bf p}^2 }
\pmatrix {{\rho\over 4} & -m & n \cr
              m  &0 & 0 \cr
              -n &0 & 0\cr }\quad +{\cal O}(\xi). \eqno (3.7)
$$
The contribution to the effective potential vanishes in the limit $\xi\to 0$,
however the amount ${1\over 2}R^\mu \Delta_{\mu\nu}R^\nu$ modifies the
B-sector and the Y-sector and provides also contributions exclusive to the
matter sector. For instance in the B-sector,
the structure of the exponent in the functional integration
is now $-{1\over 2}Q^\mu_B \Delta^{-1}_{\mu\nu}Q^\nu_B
-{1\over 2}Q^\mu_B \Theta_{\mu\nu}Q^\nu_B + Q^\mu_B\; B^\prime_\mu$ with
currents given by
$$\eqalignno {
B^\prime_0 &= B_0 + j_R { (i{\bf p}\cdot {\bf a}_Y)\over {\bf p}^2 } +
{ (i{\bf p}\cdot {\bf a}_Y)\over {\bf p}^2}
(\kappa {\bf a}_B\times {\bf Q}_Y) &(3.8)\cr
{\bf B}^\prime &= {\bf B} + (\kappa {\bf a}_Y\times {\bf {\hat z}})
{ (i{\bf p}\cdot {\bf a}_B)\over {\bf p}^2 } Q^0_Y -
{\rho\over 4\kappa {\bf p}^2} j_R {\bf a}_Y\times  {\bf {\hat z}}
+{\rho \over 4 {\bf p}^2 } ({\bf a}_B\cdot{\bf a}_Y) {\bf Q}_Y
- {\rho\over 4 {\bf p}^2} {\bf a}_B ({\bf a}_Y\cdot{\bf Q}_Y) \cr }
$$
and a matrix $\Theta$ which depends only on background fields.
Upon integrating the B-sector, one gets two contributions to the
effective potential, one that modifies the structure of the action
in the Y-sector, and one which contribute only to the matter sector.
The contributions to the effective potential are
$\ln \det^{-1/2}\Delta^{-1}_{\mu\nu}$, which vanishes in the $\xi\to 0$
limit and the second is
$\ln \det^{-1/2}\bigl ( 1 + \Delta\times \Theta \bigr )=
-{1\over 2} \ln \bigl ( 1- {({\bf p}\cdot{\bf a}_Y)^2\over {\bf p}^4}\bigr
)^2$.
The modification to the action in the Y-sector is
${1\over 2}B^{\prime\mu} \Bigl \{
\bigl [ 1 + \Delta\times \Theta \bigr]^{-1}\Delta\Bigr
\}_{\mu\nu} B^{\prime\nu}$. Upon collecting all terms that depends on the
$Q_Y^\mu$ variable, we can integrate the Y-sector in the same way.
Finally, the result of the Q-integration is divided in a contribution
to the effective potential and a part that modifies the matter
sector. We get
$$\eqalignno {
V_{\rm eff}(v, a_\mu^a) &= {1\over 2} \tr \ln \bigl (
1- {({\bf p}\cdot{\bf a}_Y)^2\over {\bf p}^4}\bigr )^2
+{1\over 2} \tr \ln \bigl ( 1- {({\bf p}\cdot{\bf a}_B)^2\over {\bf p}^4}
-{({\bf p}\cdot{\bf a}_B)^2({\bf p}\cdot{\bf a}_Y)^2\over {\bf p}^8}
\bigr )^2 \cr
&-\tr \ln \bigl ( 1- {( {\bf a}_B\cdot {\bf p})^2\over {\bf p}^4}
- {( {\bf a}_Y\cdot {\bf p})^2\over {\bf p}^4} \bigr )
+{i\over \int d^3x }\ln \int \delta\pi \exp iS_{\rm matter} &(3.9) \cr }
$$
where the first contribution in Eq.(3.9) comes from integrating the B-sector
while the second term comes from the Y-sector. The third contribution
originates
from the ghosts sector. Although the
limit $\xi \to 0$ should be taken at the end
of the calculation, we have carefully dropped terms of ${\cal O}(\xi)$ to
clarify the expressions. The modified action $S_{\rm matter}$ is given by
$$\eqalignno{
S_{\rm matter} = \int dt\;d^2{\bf x}\Bigl \{ &
i\pi^\dagger(D_t)\pi - {1\over 2}|{\bf D}\pi|^2 +{\cal L}_{{\rm S.I}}
+{\xi\over 8} (\pi^\dagger\tau^a \varphi)(\varphi^\dagger\tau^a\pi) \cr
&-{\rho \over 8\kappa^2} j_R{1\over {\bf p}^2}j_R \cr
&-{\rho \over 8\kappa^2} j_B{1\over {\bf p}^2}j_B
+ {1\over 2\kappa}(j_B-i{ {\bf p}\cdot{\bf a}_Y\over {\bf p}^2 } j_R)
{1\over {\bf p}^2}
(i{\bf p}\times {\bf a}_B) j_R \cr
&-{\rho \over 8\kappa^2} j_Y{1\over {\bf p}^2}j_Y
+{1\over 2\kappa}(j_Y-i{ {\bf p}\cdot{\bf a}_B\over {\bf p}^2 } j_R)
{1\over {\bf p}^2}
(i{\bf p}\times {\bf a}_Y) j_R \Bigr \}&(3.10) \cr }
$$
where
all expressions have the operators ${\bf p}$ and $\omega$ acting on the
right, and the covariant derivatives read
$D_t= -i(\omega-{1\over 2}a_0^R\tau^R)$ and
${\bf D}= i({\bf p}-{1\over 2}{\bf a}^a\tau^a)$.
The first line comes from the original action while the second is from
the R-integration. The third and fourth lines are from B and Y-integration
respectively.

Some comments are in order at this point. The ghosts contribution
in Eq.~(3.9) cancels against the gauge field contributions to ${\cal O}(v^4)$
leaving only
the remaining functional integration over the matter sector. Indeed, if we
had not introduced any matter fields, we would have gotten a vanishing answer
in contrast with ref. [14-17] but in agreement with [10,18,23].

In any case, when matter is present, the effective potential is
given by the remaining functional integration over the matter fields.
It is not too difficult to see that the structure of the action
(3.10) is
$$\eqalignno{
\int dt d^2{\bf x}\;\Bigl \{
& {1\over 2} \pi^{*a}_1(x) {\cal D}^{-1}_{ab}(x-x') \pi^b_1(x') +
 {1\over 2} \pi^{*a}_2(x) {\cal E}^{-1}_{ab}(x-x') \pi^b_2(x') +
J \pi^*_2 + J^*\pi_2 \Bigr \}\cr
& &(3.11) \cr}
$$ where the notation for the scalar fields is $\pi^a_i=(\pi, \pi^*)$ for
each $i=1,2$, the current mixing the $\pi$'s is given by
$J= {i\over 2} ({\bf a}_-\cdot {\bf p}) \pi_1
+{v\over 4\kappa} { ({\bf p}\times {\bf a}_-)\over {\bf p^2}} j_R$,
and the matrix for the $\pi_1$ field in Fourier space to ${\cal O}(\xi)$ is
$$
{\cal D}^{-1}(\varphi_p, a^\mu;\omega, {\bf p})
= \pmatrix { \omega - {1\over 2}{\bf p}^2 + A + {B\over {\bf p}^2}
+ {C\over {\bf p}^4}
&\bigl (-{1\over 2} \lambda -{\rho\over 16\kappa^2{\bf p}^2}
+{f \over 4\kappa{\bf p}^4}\bigr )v^2 \cr
\bigl ( -{1\over 2} \lambda -{\rho\over 16\kappa^2{\bf p}^2}
+ {f\over 4\kappa{\bf p}^4 } \bigr ) (v^*)^2
&-\omega - {1\over  2}{\bf p}^2 + A + {B\over {\bf p}^2} + {C\over {\bf p}^4}
\cr}\eqno (3.12)
$$with
${\bf a}_\pm = {\bf a}_B \pm i {\bf a}_Y$,
$A= {1\over 8}({\bf a}_+\cdot{\bf a}_-) -a^0_R/2 - \lambda\rho$,
$B= -{\rho^2\over 16\kappa^2}$, $C={\rho\over 4\kappa}f$ and
$f=-[({\bf p}\cdot{\bf a}_B)({\bf p}\times{\bf a}_Y) +
({\bf p}\cdot{\bf a}_Y)({\bf p}\times{\bf a}_B)]$.

Similarly, the matrix for the $\pi_2$ field is
$$
{\cal E}^{-1}(\varphi_p, a^\mu;\omega, {\bf p})
= \pmatrix { \omega - {1\over 2}{\bf p}^2 + E + {F\over {\bf p}^2}
&0\cr
0
&-\omega - {1\over  2}{\bf p}^2 + E + {F \over {\bf p}^2}    \cr
}\eqno (3.13)
$$
with $E= {1\over 8}({\bf a}_+\cdot{\bf a}_-) + a^0_R/2 - {1\over
2}\lambda\rho$,
and $F= -{1\over 8}{\rho^2\over \kappa^2}$. To perform
the functional integration
over $\pi_2$ is not difficult.
Upon doing it, there remains only to perform the integration over
$\pi_1$ and the
result, keeping in mind that we are computing up to
include ${\cal O}(v^4)$ in the
limit $\xi \to 0$, is
$$\eqalignno {
V_{\rm eff} (v, a_\mu^a)\int d^3x &=
i\ln \int \delta\pi \exp iS_{\rm matter}
 =-{i\over 2}\ln {\rm Det} {\cal E}^{-1}_{ab}
-{i\over 2}\ln {\rm Det} \Bigl \{ {\cal D}^{-1}_{ab} + {\cal M}_{ab}\Bigr \}
\quad
&(3.14) \cr }
$$
where the easily found ${\cal M}_{ab}$ matrix
appears as a consequence of the mixing between the
$\pi$'s and the determinant are taken functionally. Since
the operators ${\cal E}^{-1}$ and ${\cal D}^{-1}$ are
diagonal in Fourier space,
we can write the operatorial form of the
final contribution to the effective potential to ${\cal O}(v^4)$ and in the
limit $\xi \to 0$ as
$$\eqalignno {
V_{\rm eff} (v, a_\mu^a)=&
-{i\over 2} \tr \ln {1\over \mu^{\prime 4}}
\Bigl \{ -\omega^2 + ({\bf p}^2/2-E)^2 -F \Bigr \} \cr
&-{i\over 2} \tr \ln {1\over \mu^{\prime 4}}
\Bigl \{ -\omega^2 + ({\bf p}^2/2- A - {B\over {\bf p^2}} -{C\over {\bf
p^2}})^2
-{1\over 4}\lambda^2 \rho^2 \cr
& \quad\quad\quad +\omega (X_+ - X_-) + (A - {{\bf p^2}\over 2})
(X_+ + X_-) + X_+ X_- \Bigr \}
&(3.15) \cr}
$$
where the trace is performed in energy/momentum space,
the parameter $\mu^\prime$ of mass dimension one is introduced for dimensional
reasons [26], and
$$
\eqalignno{
X_\pm =& {1\over 4}\Delta^{-1}_E\Bigl \{
({\bf p}\cdot {\bf a_+}) (\pm\omega-{\bf p^2}/2 + E)({\bf p}\cdot {\bf a_-})
\cr
&-i{\rho\over 4\kappa} ({\bf p}\times{\bf a}_+) (\pm\omega-{\bf p^2}/2)
({\bf p}\cdot {\bf a_-})
+i {\rho\over 4\kappa} ({\bf p}\times{\bf a}_-) (\pm\omega-{\bf p^2}/2)
({\bf p}\cdot {\bf a_+}) \Bigr \}
&(3.16)\cr }
$$
with $\Delta_E\equiv -\omega^2+({\bf p^2}/2-E)^2$.

We pose for a moment to note that so far we have not used any form of
regulator to extract the information we have in Eq.(3.15). This is because
we have not encountered any ultraviolet divergences so far. However, Eq.(3.15)
is divergent in the ultraviolet regime
and therefore requires a regulator in order
to evaluate its contribution to the effective potential.
We will use operator regularization [22,26,27] to perform the
computation since it preserve all symmetries present at the classical
level modulo anomalies.

For each logarithm in Eq.(3.15), it is necessary to
identify an operator $H_0$ and an operator $H_I$. Upon using operator
regularization, the n-point function is easily identified as the
n-th
$H_I$ insertion with $H_0$ acting as the propagator for each internal lines.
Following Ref.[22], we define
$H_0=\{ -\omega^2 + ({\bf p}^2/2-A_1)^2\}/\mu^{\prime 4}$
for the first logarithm and
$H_0=\{ -\omega^2 + ({\bf p}^2/2-E_1)^2\}/\mu^{\prime 4}$
for the second one where
$A=A_1+A_2$, $E=E_1+E_2$, $A_1=-\lambda\rho$ and
$E_1=-{1\over 2}\lambda\rho$. In
$H_I$, we collect the rest of the expressions for each logarithm.

Both logarithm are easily regulated via
$$
{\rm Tr} \ln H=-\lim_{s\to0}{d \over ds}{\rm Tr}
{1\over \Gamma(s)}\int_0^\infty dt\; t^{s-1}
\Bigl\{e^{-H_0t}+e^{-H_0t}(-t)H_I+e^{-H_0t}{(-t)^2\over 2}H_I^2+\dots\Bigr\}
$$
and upon using the useful integral over the energy $d\omega$
$$\ick {
I\equiv& \int_{-\infty}^{\infty}{d\omega\over 2\pi}{1\over \Delta_a^{1+s}}
=i{(2s)!\over s!s!}({\bf p}^2+2a)^{-(1+2s)}\quad ,&(3.17) \cr }
$$
the contribution to the effective potential from the first log is
$$
\ick{
&-{i\over 2}\int {d^2{\bf p}\over (2\pi)^2}\Bigl ({d\over ds}s\Bigr )
\Bigl\{[2E_1E_2+E_2^2-F]-{\bf p^2}E_2\Bigr\}i{(2s)!\over s!s!}
{(\mu^{\prime})^{4s}\over
({\bf p}^2-2E_1)^{1+2s}} \cr
&+{i\over 4}\int {d^2{\bf p}\over (2\pi)^2}\Bigl ({d\over ds}s(s+1)\Bigr )
{\bf p}^4E_2^2 i{(2s+2)!\over (s+1)!(s+1)!}
{(\mu^{\prime})^{4s}\over ({\bf p}^2-2E_1)^{3+2s}} &(3.18)\cr }
$$
where the first term is a one-pts function while the second is a two-pts
function. Upon
symmetric integration over momentum integrals, (3.18) becomes
$$\ick {
&-{1\over 8\pi}F\ln\Bigl({\mu^{\prime 2}\over -2E_1}\Bigr) \quad .
&(3.19) \cr }
$$
The second logarithm is more tedious to compute as it involves many one and
two-pts functions. We rewrite the second logarithm of Eq.(3.15) as
$$
\ick{
-{i\over 2}\int d\omega d^2{\bf p} \ln{1\over \mu^{\prime 4} } \Bigl\{
&-\omega^2+\Bigl({{\bf p}^2\over 2}-A_1\Bigr)^2-A_2{\bf p}^2
+2A_1A_2+A_2^2-{1\over 4}\lambda^2\rho^2-B-{C\over {\bf p}^2} \cr
&-{i\over 2\kappa} \Bigl( \omega^2 + {{\bf p}^4\over 4} \Bigr)
{\rho\over 4{\bf p}^2} [({\bf p}\times{\bf a}_+)({\bf p}\cdot{\bf a}_-)-
({\bf p}\times{\bf a}_-)({\bf p}\cdot{\bf a}_+)]{1\over \Delta_{E_1}}\cr
&+\Bigl(\omega^2+{{\bf p}^4\over 4}\Bigr)
{1\over 2}({\bf p}\cdot{\bf a}_+)({\bf p}\cdot{\bf a}_-)
{{\bf p}^2E_2\over \Delta_{E_1}^2}-{{\bf p}^2\over 4}(A+E)
{({\bf p}\cdot {\bf a}_+)({\bf p}\cdot {\bf a}_-)\over \Delta_{E_1}^2} \cr
&+\Bigl(\omega^2+{{\bf p}^4\over 4}\Bigr){1\over 2}
{({\bf p}\cdot {\bf a}_+)({\bf p}\cdot {\bf a}_-)\over \Delta_{E_1}}
+{1\over 16}{({\bf p}\cdot {\bf a}_+)^2({\bf p}\cdot
{\bf a}_-)^2\over \Delta_{E_1}}
\Bigr\} &(3.20) \cr}
$$
Upon using the regulated form of the logarithm, Eq.~(3.17)
and symmetric integration over
$d^2{\bf p}$, we obtain for the non-vanishing one-pts functions
$$
\ick{
-{1\over 32\pi}\Bigl\{&8A_1A_2 - 4[2A_1A_2+A_2^2-{1\over 4}\lambda^2\rho^2-B]
-E_2({\bf a}_+\cdot{\bf a}_-)  \cr
&+(A+E)({\bf a}_+\cdot{\bf a}_-)
-(A_1+E_1)({\bf a}_+\cdot{\bf a}_-)
-{1\over 16}{\cal P}\Bigr\}\ln{\mu^{\prime 2}\over -2A_1} &(3.21) \cr}
$$
where ${\cal P}=\bigl\{({\bf a}_+\cdot{\bf a}_+)({\bf a}_-\cdot{\bf
a}_-)+2({\bf
a}_+\cdot{\bf a}_-)^2\bigr\}$ and from the non-vanishing two-pts function
$$\ick {
&-{1\over 32\pi}\Bigl\{4A_2^2-A_2({\bf a}_+\cdot{\bf a}_-)+{1\over 16}{\cal P}
\Bigr\}\ln{\mu^{\prime 2}\over -2A_1}&(3.22) \cr }
$$
where each terms in Eq.~(3.22) arises separately from squaring the
$A_2{\bf p}^2$-term of Eq.~(3.20),
from crossing the $A_2{\bf p}^2$-term with the one before last
of Eq.~(3.20), and
from squaring the one before last of Eq.~(3.20), respectively.

Upon collecting all contributions of Eq.~(3.19,21,22), we obtain for the
unnormalized effective potential
$$
\ick{
V_{\rm eff}(\rho,a_\mu^a)
&={1\over 4}\lambda\rho^2 + c_1\rho^2 -
{1\over 8\pi}\Bigl (F+({1\over 4}\lambda^2\rho^2+B)\Bigr )
\ln{\mu^{\prime 2}\over -2A_1}\cr
&={1\over 4}\lambda\rho^2+c_2\rho^2+{1\over 8\pi}\Bigl (4\lambda^2-{3\over
\kappa^2}\Bigr ){\rho^2\over 16}\ln{\rho \over \mu^{\prime 2} } &(3.23) \cr }
$$
where now global gauge invariance is restored with
$\rho=v_p^{\ast}v_p$.
In obtaining Eq.~(3.19) and Eqs.~(3.21-22), we drop
an unimportant (const.$\rho^2$)-term
arising from the first term independent of
$H_I$ in the regulated form of the logarithm. We have inserted this
contribution
in the $c_1\rho^2$-term in Eq.~(3.23) together with a term of the same form
which arises from Eq.~(3.19) because $A_1=2E_1$. The $c_2\rho^2$-term collects
the $c_1\rho^2$-term with the term proportional to $\rho^2\ln 2\lambda$.
In any case, the $c_2\rho^2$-term disappear upon normalizing the effective
potential. Note that no ultraviolet divergences occur in Eq.~(3.23)
as expected upon using operator regularization.

After imposing the normalization condition
$$
{d^2\over d\rho^2}V_{\rm eff}|_{\rho=\mu^2}={1\over 2}\lambda(\mu),\eqno (3.24)
$$
the normalized effective scalar field potential in the ${\rm R}_\xi$-gauge in
$\xi\to 0$ limit up to include ${\cal O}(v^4)$ contributions is
$$\ick {
V_{\rm eff}(\rho,a_\mu^a)&={1\over 4}\rho^2\Bigl[\lambda(\mu)+{1\over
8\pi}\bigl(\lambda^2(\mu)-{4\over \kappa^2}\alpha^2 \bigr)
\bigl( \ln{\rho \over\mu^2} - {3\over 2}\bigr )\Bigr].&(3.25)\cr }
$$
where the appearance of the group theoretical factor $\alpha^2=3/16$ is a
consequence of the su(2) Lie algebra: 3 corresponds to the number
of generators and 1/16 to a normalization of the generators.
Note that the background gauge fields do not contribute to the
scalar field effective potential.

\vglue 0.6cm
\line{\bf 4. Summary and conclusions \hfil}
\vglue 0.4cm

We computed the scalar field effective potential of a nonrelativistic
non-Abelian Chern-Simons field theory possessing various classical symmetries
such as Galilean, conformal and gauge symmetries. We applied the traditional
functional method using constant background gauge and matter fields
in order to satisfy Gauss's law. Simplifications in the course of the
calculation are manifest
when constant background gauge and matter fields are used
since the determinants are taken on $3\times 3$ constant matrix,
which are diagonal in Fourier space, and when an $R_\xi$ gauge-fixing condition
is imposed, which respect Galilean invariance and global gauge invariance.
We have regulated the divergences in the matter sector using operator
regularization. We note that the scalar field effective
potential does not depend, to the order considered, on the background gauge
field, which satisfies Gauss's law and that our result is in agreement with
a diagrammatic analysis of the non-Abelian Aharonov-Bohm scattering. As a spin
off of our calculation, we find that there are no infinite nor finite
renormalization of the Chern-Simons coupling constant $\kappa$ in our method in
contrast to the results of ref.[14-17] but in agreement with [10,18,23].

We note that the effective potential presented in Eq.~(3.25) is
a generalization of the effective potential found in the Abelian version
of the model (1.1), which can be retrieved by setting $\alpha^2=1$ [22].

We did not discuss here the gauge parameter dependence of our result
Eq.~(3.25). However,
in the Abelian version of the model (1.1), the effective potential was also
computed with the $R_\xi$ gauge-fixing condition and with a Coulomb
gauge with arbitrary $\xi$. We found in that case, that the effective potential
was the same in either gauge-fixing conditions and was independent
of the gauge parameter $\xi$ [22]. We therefore expect that our result for the
effective potential presented in Eq.~(3.25) to be gauge parameter independent.

Finally, we analyse the scale anomaly. Conformal symmetry is related to the
$\beta$-function. A non-vanishing $\beta$-function indicates conformal symmetry
breaking. Using the renormalization group equation
$$
0=\mu{d\over d\mu}V_{\rm eff}(\rho)=\Bigl [ \mu {\partial\over{\partial\mu}}
+\beta(\lambda_1(\mu)){\partial\over {\partial\lambda_1(\mu)}}
\Bigr ] V_{\rm eff} (\rho) \eqno(4.1)
$$
the $\beta$-function reads
$$
\beta(\lambda(\mu))={1 \over 4\pi}\Bigl (\lambda^2(\mu) -
{4 \over \kappa^2}{3\over 16} \Bigr )\quad .\eqno(4.2)
$$
For unrelated coupling constants the theory loses conformal symmetry.
At the self-dual point
$\lambda(\mu)=-{{\sqrt 3}\over 2\kappa}$ and at
$\lambda(\mu)={{\sqrt 3}\over 2\kappa}$ the
$\beta$-function vanishes; hence, the theory is conformally symmetric,
recovering the result of Bak and Bergman [23].

\vfill\eject
\vglue 0.6cm
\line{\bf Acknowledgements \hfil}
\vglue 0.4cm

We thank R.B. MacKenzie, D.G.C. McKeon and M.B. Paranjape for useful comments.

\vglue 0.6cm
\line{\bf References \hfil}
\vglue 0.4cm

\item{1.} J. Hong, Y. Kim and P.Y. Pac, Phys. Rev. Lett. {\bf 64}, 2230 (1990).
\item{2.} R. Jackiw and E.J. Weinberg, Phys. Rev. Lett. {\bf 64}, 2234 (1990).
\item{3.} See for a review, J. D. Lykken, J. Sonnenschein and N. Weiss,
Int. J. of Mod. Phys. {\bf A6}, 5155 (1991).
\item{4.} C. Lee, K. Lee and E.J. Weinberg, Phys. Lett. {\bf B243}, 105 (1990).
\item{5.} M. Leblanc and M.T. Thomaz, Phys. Lett. {\bf B281}, 259 (1992).
\item{6.} C.R. Hagen, Phys. Rev. {\bf D31}, 848 (1985).
\item{7.} R. Jackiw and S.Y. Pi, Phys. Rev. {\bf D49}, 3500 (1990).
\item{8.} G. Lozano, M. Leblanc and H. Min, Ann. of Phys.{\bf 219}, 328 (1992).
\item{9.} S. Deser, R. Jackiw and S. Templeton, Phys. Rev. Lett.
{\bf 48}, 975 (1982); Ann. Phys. (N.Y.) {\bf 140}, 372 (1982).
\item {10.} W. Chen, G.W. Semenoff and Y. Wu, Phys. Rev. {\bf D 46},
5521 (1992) [and references therein].
\item {11.} G. Dunne, R. Jackiw, S. Pi and C. Trugenberger,
Phys. Rev. {\bf D 43}, 1332 (1991).
\item {12.} E.A. Ivanov, Phys. Lett. {\bf 268B}, 203 (1991);
S.J. Gates and H. Nishino, {\it ibid.}{\bf 281}, 72 (1991).
\item {13.} K. Lee, Phys. Rev. Lett. {\bf 66}, 553 (1991); K. Lee, Phys. Lett.
{\bf 255B}, 381 (1991).
\item {14.} R. Pisarski and S. Rao, Phys. Rev. {\bf D32}, 2081 (1985).
\item {15.} E. Witten, Commun. Math. Phys. {\bf 121}, 351 (1989).
\item {16.} L. Alvarez-Gaum\'e, J.M.F. Labastida and A.V. Ramallo,
Nucl. Phys. {\bf B334},103 (1990).
\item {17.} D. Birmingham, R. Kantowski and M. Rakowski, Phys. Lett. {\bf
B251},
121 (1990).
\item {18.} G. Giavarini, C.P. Martin and F. Riuz Riuz, preprint
FTUAM94/8, NIKHEF-H 94/14, UPRF94/395, hepth/9406034.
\item {19.} G. Lozano, Phys. Lett. {\bf B283}, 70 (1992).
\item {20.} O. Bergman and G. Lozano, Ann. Phys. (N.Y.) {\bf 229}, 416 (1994).
\item {21.} D. Freedman, G. Lozano and N. Rius, Phys. Rev. {\bf D49}, 1054
(1994).
\item {22.} D. Caenepeel, F. Gingras, M. Leblanc and D.G.C. McKeon,
Phys. Rev. {\bf D49}, (1994) (in press).
\item {23.} D. Bak and O. Bergman, preprint MIT-CTP-2283, (1994).
\item {24.} R. Jackiw, Ann. Phys. (N.Y.) {\bf 201}, 83 (1990).
\item {25.} R. Jackiw, Phys. Rev. {\bf D9}, 1686 (1974).
\item {26.} D.G.C. McKeon, T.N. Sherry, Phys. Rev. {\bf 59}, 532 (1987); Phys.
Rev. {\bf D35}, 3584 (1987); Can. J. Phys. {\bf 66}, 268 (1988).
\item {27.} L. Culumovic, M. Leblanc, R.B. Mann, D.G.C. McKeon, and T.N.
Sherry,
Phys. Rev. {\bf D41}, 514 (1990).
\item {28.} L.S. Brown and W.I. Weisberger, Nucl. Phys. {\bf B157}, 285 (1979).
\item {29.} L.F. Abbott, Nucl. Phys. {\bf B185}, 189 (1981).

\bye